\documentclass[prl,twocolumn]{revtex4}
\usepackage{array,amsmath,amssymb,graphicx}
\begin{document}

\title{Low-frequency dynamics of disordered $XY$ spin chains and pinned
density waves: from localized spin waves to soliton tunneling}

\author{Michael M. Fogler}


\affiliation{Department of Physics, Massachusetts Institute of
Technology, 77 Massachusetts Avenue, Cambridge, MA 02139}

\date{\today}

\begin{abstract}

A long-standing problem of the low-energy dynamics of a disordered $XY$
spin chain is re-examined. The case of a rigid chain is studied where
the quantum effects can be treated quasiclassically. It is shown that as
the frequency decreases, the relevant excitations change from localized
spin waves to two-level systems to soliton-antisoliton pairs. The
linear-response correlation functions are calculated. The results
apply to other periodic glassy systems such as pinned density
waves, planar vortex lattices, stripes, and disordered
Luttinger liquids.

\end{abstract}

\maketitle

In this Letter I revisit the problem of the low-frequency response of a
periodic elastic string pinned by a weak random potential. This problem
has been repeatedly studied in connection with dynamics of density
waves, Luttinger liquids, glasses, and random-field $XY$ spin chains. It
is also germane for statistical mechanics of periodic planar systems
(Josephson junctions, stripes) with columnar disorder. The approach
presented below is customized to one dimension (1D) but the results
should apply to all dimensions $d < 4$ after a few minor modifications.
The difficulty of the problem stems from strong nonperturbative disorder
effects at low enough frequencies and large enough length scales. Such
effects are termed {\it pinning\/}~\cite{Blatter_94} or {\it
localization\/} depending on a context. Rigorous results regarding the
pinned regime are restricted to one particular value of the chain
elastic modulus where the problem maps onto noninteracting fermions in a
random potential, whose dynamics is governed by the famous
Mott--Halperin--Berezinskii (MHB) law~\cite{Mott_68,Berezinskii_73}. In
this special case the chain is extremely soft and behaves as a quantum
object. On the other hand, real charge density waves (CDW) and
low-density electron gases are much more rigid and consequently
quasiclassical. Despite many previous
attempts~\cite{Fukuyama_77,Feigelman_80,Giamarchi_96} to calculate the
response of the latter type of systems, no fully satisfactory solution
has emerged. The existing controversy between different authors is
clouded by an uncontrolled nature of approximations they use. The purpose
of this Letter is to demonstrate a feasibility of a quantitative
analysis of the problem and to elucidate the physical picture of the
low-frequency excitations in the glassy phase of a rigid (strongly
interacting) system. Most of the discussion is focused on the classical
dynamics; however, at the end I also consider quantum effects and
establish the connection with the MHB law. In plain words, I offer a
definitive answer the following basic question: {\it If we shake a
pinned elastic chain, how does it respond\/}?

{\it Model\/}.--- Consider an $XY$ spin chain
with spin stiffness $c$, spin wave
velocity $v$, lattice constant $a$, which is
described by the Lagrangean
\begin{eqnarray}
{\cal L} = \sum\limits_j \biggl\{
\frac{c}{2}\biggl[\frac{a}{v^2} (\partial_t \phi_j)^2 &-&
\frac{v^2}{a} [\phi(x_{j + 1}) - \phi(x_j)]^2\biggr]
\nonumber\\
\mbox{} - h_j e^{i \phi(x_j)} &-& h_j^* e^{-i \phi(x_j)}
\biggr\},
\label{L}
\end{eqnarray}
where $x_j = a j$ and $h_j$ are quenched Gaussian random variables with
zero mean and variance $\langle |h_j|^2 \rangle = \Delta$. Depending on
the context, $\phi$ can equally well represent the phase of a CDW, the
bosonized density field of a Luttinger liquid or a transverse
displacement of a flexible line object. The quantity of interest is the
propagator $D(\omega, q)$ of the field $\phi$, or equivalently, the
quantity
\begin{equation}
\sigma(\omega, q) = -i \frac{e^2 \omega}{4 \pi^2 \hbar^2} D(\omega, q),
\label{sigma}
\end{equation}
which has the meaning of the ac conductivity.
Of particular interest is $\sigma(\omega, q = 0) \equiv \sigma(\omega)$.

{\it Classical glass\/}.--- A classical chain respond to an external
perturbation by elastic vibrations around the ground state $\phi^0(x)$.
It is therefore instructive to recall the basic properties of $\phi^0$,
described by the collective pinning
theory~\cite{Fukuyama_77,Blatter_94}. The spatial structure of $\phi^0$
is determined by the competition between disorder and elasticity: on
short scales the elasticity prevents large phase distortions but on long
scales the disorder eventually wins and the distortions grow without a
bound: $\langle [\phi^0(x) - \phi^0(0)]^2 \rangle \to \infty$ as $x \to
\infty$. There is a characteristic scale $R_c$ where a $2\pi$-distortion
is accumulated. At this scale the typical elastic and disorder energies
of a chain segment are of the same order, $c / R_c \sim \sqrt{R_c \Delta
/ a}$. It is easy to see that $R_c \sim (c^2 a / \Delta)^{1/3}$. A crude
but useful picture~\cite{Fukuyama_77} is to imagine that the chain
consists of domains of size $R_c$ individually pinned by collective
potential wells generated by random fields $h_j$.

An explicit algorithm for finding $\phi^0(x)$ was originally given by
Feigelman~\cite{Feigelman_80}. Consider the segment of the chain with
$j$ leftmost spins, which satisfies the boundary conditions at the left
and has the angular coordinate of the $j$th spin fixed at a given value
$\phi_j$. Let $E^-(x_j, \phi_j)$ be the minimal energy of such a segment
optimized with respect to all spin coordinates in the interior of the
segment. Function $E^-$ satisfies the recurrence
relation~\cite{Feigelman_80}
%
\begin{eqnarray}
E^-(x_{j + 1}, \phi) &=& \min_{\phi^\prime} \left\{ E^-(x_j, \phi^\prime)
 + \frac{c}{2 a} (\phi - \phi^\prime)^2
\right\}
\nonumber\\
 &+& a U(x_{j + 1}, \phi),
\label{discrete_KPZ}\\
U(x_{j + 1}, \phi) &\equiv& a^{-1} (h_{j + 1} e^{i \phi} + h_{j + 1}^* e^{-i \phi}).
\label{U}
\end{eqnarray}
Let $E^+(x_j, \phi)$ be a similar function for the right end of the
chain, then the desired $\phi^0(x_{j})$ is the value of $\phi$ that
minimizes the sum $E(x_{j}, \phi) = E^-(x_{j}, \phi) + E^+(x_{j},
\phi)$. For $R_c \gg a$, $\phi^0$ varies little from $j$ to $j + 1$, and
so Eq.~(\ref{discrete_KPZ}) possesses a meaningful continuum limit,
\begin{equation}
E^-_x = -(1 / 2 c) (E^-_\phi)^2 + U(x, \phi),
\label{KPZ}
\end{equation}
which is the Kardar-Parisi-Zhang (KPZ) equation~\cite{Feigelman_80,Kardar}.

\begin{figure}
\centerline{
\includegraphics[width=1.6in,bb=143 430 436 649]{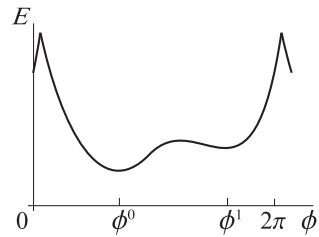}
}
\setlength{\columnwidth}{3.2in}
\caption{Typical behavior of $E(\phi)$ (for a given fixed $x$).
\label{Fig_E}
}
\end{figure}

Due to the $2\pi$-periodicity of $U$ the solutions of
Eqs.~(\ref{discrete_KPZ}) and~(\ref{KPZ}) become $2\pi$-periodic at
large $x$ irrespectively of the boundary conditions at the ends. A
typical behavior of $E(\phi)$ is illustrated in Fig.~\ref{Fig_E}.
$E(\phi)$ gives direct information about the rigidity of the system. The
minimal work needed to twist a given spin of the chain to the angular
coordinate $\phi$ is equal to $E(\phi) - E(\phi^0)$. For small
$\delta\phi = \phi - \phi^0$, it is quadratic in $\delta\phi$. The
elastic distortion caused by the twist is localized predominantly within
a domain of length $R_c$ around the chosen spin. When $|\delta\phi|$
exceeds a certain value, the chain suddenly snaps into a conformation
corresponding to a competing metastable state. At such $\phi$, function
$E(\phi)$ exhibits upward cusps (typically, one per period --- see
Fig.~\ref{Fig_E}), extensively studied in the context of the KPZ
equation and Burgers turbulence~\cite{Kardar,E_97}.

Of interest to us here are the low-frequency excitations of the chain.
They can be visualized as localized mechanical oscillations of
essentially rigid segments of size $R_c$. This concept can be succinctly
expressed by means of the following low-frequency local effective action
\begin{equation}
{\cal L}_{\rm eff} = \frac12 M (\partial_t \phi)^2 - E(\phi),
\label{L_eff}
\end{equation}
which describes the motion of a particle (``domain'') of
mass $M \sim R_c c / v^2$ in the potential $E(\phi)$. To justify this
action, I will follow Ref.~\cite{Feigelman_80} but
with important modifications, leading to very different end results.

To calculate the linear response we expand ${\cal L}$ [Eq.~(\ref{L})] in
$\delta\phi$ and keep only the second order terms. After that we perform
the usual spectral decomposition to obtain
\begin{equation}
D(\omega; x, x^\prime) = \frac{1}{a}
\sum\limits_n
\frac{\psi_n(x) \psi_n(x^\prime)}{\varepsilon_n - \varepsilon(\omega) + i 0},
\quad \varepsilon(\omega) \equiv \frac{c \omega^2}{v^2}.
\label{D_spectral_expansion}
\end{equation}
The frequency $\omega_n$ of $n$th eigenmode and its wavefunction
$\psi_n$ satisfy the discrete Schr\"oedinger equation
\begin{equation}
-\frac{c}{2} \nabla^2 \psi_n(x_j)
+ U(x_j, \phi^0(x_j)) \psi_n(x_j) = \varepsilon_n \psi_n,
\label{Schroedinger}
\end{equation}
where $\nabla$ is the lattice derivative and $\varepsilon_n =
\varepsilon(\omega_n)$ is the ``energy'' of the mode. From the scaling
of the ``kinetic'' and the disorder energies with distance, we quickly
deduce that the low-energy modes are necessarily trapped in the
potential wells of the random potential $U$. To determine their typical
localization length $L_{\rm loc}$ we can compare the kinetic energy $c /
L_{\rm loc}^2$ of a wavepacket of size $L_{\rm loc}$ with the depth
$\sqrt{\Delta / L_{\rm loc} a}$ of a typical potential well, which
yields $L_{\rm loc} \sim R_c$. Thus, $R_c$ is the unique characteristic
length of the classical glass regime, which shows up both in statics and
dynamics~\cite{Comment_on_wavefunctions}. The crossover to this regime
occurs around the {\it pinning frequency\/} $\omega_p = v / R_c$. Note
that the depths of individual potential wells have a broad distribution
around the typical value $\sqrt{\Delta / R_c a} \sim
\varepsilon(\omega_p)$, which gives rise to an inhomogeneously broadened
spectrum extending down to $\varepsilon = 0$. Modes with $\varepsilon_n
\ll \varepsilon(\omega_p)$ stem from near cancellations between the
kinetic and potential terms.

%
%

Let $x$ be a point near which one of such modes, $\psi_n$, is localized,
then for all $\omega < \omega_p$, Eq.~(\ref{D_spectral_expansion})
implies that $D(\omega; x, x) = A / (\varepsilon_n - \varepsilon + i 0)
+ {\rm smaller\ terms}$, with $A \equiv \psi_n^2(x) / a \sim R_c^{-1}$.
On the other hand, $[D(0; x, x)]^{-1} = E_{\phi \phi}(\phi^0) \equiv
\alpha(x)$~\cite{Feigelman_80}, and so $[D(\omega; x, x)]^{-1} = \alpha
- M \omega^2 + o(\alpha)$. This indicates that ${\cal L}_{\rm eff}$
indeed has the correct form for small $\delta\phi$. (This is {\it all\/}
we need to start using ${\cal L}_{\rm eff}$ for calculating the
classical linear response). Equation~(\ref{L_eff}) is also correct for
adiabatically slow motions, $\partial_t \phi \to 0$. Thus, it cannot be
badly wrong for all $\delta\phi$ smaller than the distance from $\phi_0$
to the nearest cusp of $E(\phi)$. Indeed, local smooth distortions
should appear to the rest of the system as adiabatically slow ones
provided their characteristic frequencies are sufficiently small,
$\omega \ll \omega_p$.

From Eq.~(\ref{L_eff}) we see that the disorder-averaged low-$\omega$
response of the chain is encoded in the small-$\alpha$ behavior of the
distribution function $P(\alpha)$ of the local ground-state rigidity
$\alpha(x)$, e.g., the spin wave density of states $\rho(\varepsilon) =
(N_s a)^{-1} \sum_n \delta(\varepsilon - \varepsilon_n)$ is given by
\begin{equation}
 \rho(\varepsilon) \sim P(\alpha = \varepsilon R_c),\quad
 \varepsilon < \varepsilon(\omega_p).
\label{P_and_rho}
\end{equation}
Here $N_s$ is the number of spins in the chain. Similarly, starting from
the spectral representation
\begin{equation}
 {\Re\rm e}\,\sigma(\omega) = \frac{e^2 \omega}{4 \pi N_s}
\textstyle 
\left\langle{\sum_n} d_n^2 \delta(\varepsilon - \varepsilon_n)
\right\rangle,
\label{sigma_spectral_expansion}
\end{equation}
where $d_n \equiv {\sum_j} \psi_n(x_j)$, we obtain $d_n \sim (R_c /
a)^{1/2}$ and
\begin{equation}
{\Re\rm e}\,\sigma(\omega) \sim e^2 \omega R_c \rho[\varepsilon(\omega)],
\quad \omega < \omega_p.
\label{sigma_and_rho}
\end{equation}

Now I intend to show that at small $\alpha$
\begin{equation}
               P(\alpha) \propto \alpha^s,
\label{P_alpha}
\end{equation}
where $s = 3 / 2$. Combined with Eqs.~(\ref{P_and_rho}) and
(\ref{sigma_and_rho}), this entails~\cite{Comment_on_s} 
\begin{equation}
{\Re\rm e}\,\sigma(\omega) \sim (e^2 v R_c / c)
	({\omega}/{\omega_p})^4,\quad \omega \ll \omega_p.
\label{classical_glass}
\end{equation}
Recall that $\alpha$ is defined to be the second derivative of $E(\phi)$
at the point of its global minimum $\phi_0$. Let us first demonstrate
that at arbitrary {\it local\/} minima ${\rm Prob}(\alpha) \equiv
P_l(\alpha) \propto |\alpha|$ at small $\alpha$. Indeed, local extrema
are the points where $E_\phi(\phi) = 0$; hence, $P_l(\alpha) = \langle
N_m^{-1} \int_0^{2 \pi} d \phi \delta(E_{\phi \phi} - \alpha)
\delta(E_\phi) |E_{\phi \phi}| \rangle$ where $N_m = \int_0^{2 \pi} d
\phi \delta(E_\phi) |E_{\phi \phi}|$. (The term $|E_{\phi \phi}|$ in the
integrands is the Jacobian). In the limit $\alpha \to 0$ we obtain
\begin{equation}
\textstyle  P_l(\alpha) = |\alpha| \left\langle N_m^{-1} \int\limits_0^{2 \pi}
  d \phi \delta(E_{\phi \phi}) \delta(E_\phi)\right\rangle
            = C_0 |\alpha|.
\label{P_alpha_local_II}
\end{equation}
The main contribution to the integral that determines the constant $C_0
\sim R_c^2 / c^2$ is supplied by configurations where $E^-_{\phi \phi}$
and $E^+_{\phi \phi}$ have their typical values of the order of $c /
R_c$ but are opposite in sign and almost cancel each other. Thus, soft
modes arise from the frustrations in the ground state, e.g., when $\phi
= \phi_0$ gives a rather low energy to the left half of the chain but
corresponds to a high energy state of the right half. Such frustrations
can always happen because $E^-$ and $E^+$ are uncorrelated.

%
%
\begin{figure}
\centerline{
\includegraphics[width=2.0in,angle=90,bb=20 66 520 765]{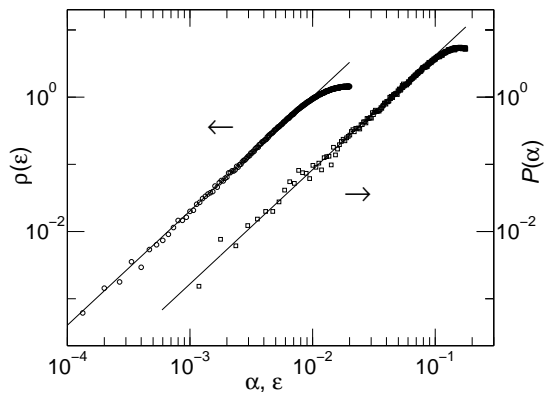}
}
\setlength{\columnwidth}{3.2in}
\caption{
Symbols: $\rho(\varepsilon)$ and $P(\alpha)$ computed and averaged over
$10^6$ disorder realizations numerically. Simulation parameters: $c =
1$, $N_s = 200$, $\Delta = 7\times 10^{-7}$. Taking previous definitions
{\it literally,\/} we get $R_c \sim 112 a$. The actual correlation length of
$\phi^0$ is about $10 a$. Straight lines: fits to the power laws with
exponent $s = 1.7$. 
\label{Fig_DOS}
}
\end{figure}

Let us now consider the {\it global\/} minimum. In its vicinity
\begin{equation}
E(\phi) = E(\phi^0) + \frac{\alpha}{2} \delta\phi^2 + \beta \delta\phi^3
        + \gamma \delta\phi^4,
\label{E_expansion}
\end{equation}
which leads to the double-minimum structure depicted in
Fig.~\ref{Fig_E}. It is easy to see that $\phi^0$ is the lower of the
two minima only if $|\beta| < \sqrt{2 \alpha \gamma}$. Via a
straightforward extension of the argument leading to
Eq.~(\ref{P_alpha_local_II}), one can obtain $P_l(\alpha,\beta) = C_1
|\alpha|$ for the {\it joint\/} distribution function of small $\alpha$
and $\beta$. The restriction on allowed $\beta$ at the global minimum
leads to $P(\alpha) \propto \sqrt{\alpha} P_l(\alpha, \beta) \propto
\alpha^{3/2}$ as I claimed above. Here I ignore competing minima away
from the immediate vicinity of $\phi^0$ because their number within a $2
\pi$-period is $N_m \sim 1$. They cannot yield any additional powers of
the small parameter $\alpha$.

Now I present my numerical results. The numerical solution of
Eq.~(\ref{discrete_KPZ}) does not pose major difficulties. This has to
be done for a number of disorder realizations $\{h_j\}$, followed by
solving the eigenvalue problem~(\ref{Schroedinger}), calculating the
desired response functions, and their statistical averaging. Such a
procedure produces the rigidity distribution $P(\alpha)$ shown in
Fig.~\ref{Fig_DOS}. The power-law behavior predicted by
Eq.~(\ref{P_alpha}) is apparent. Function $\rho(\varepsilon)$, plotted
in the same figure, can be well fitted to the power law with the same
exponent, which shows that Eq.~(\ref{P_and_rho}) is also well satisfied.
The actual value of the exponent from those fits, $s = 1.7$, deviates
slightly from my analytical prediction $s = 3/2$. Further numerical work
is needed to resolve this small discrepancy but there is enough accuracy
to rule out $s = 1/2$~\cite{Fukuyama_77,Giamarchi_96} or $s =
1$~\cite{Feigelman_80}. As for the functional shape of $E(\phi)$, I
observed the double-minimum structure sketched in Fig.~\ref{Fig_E} quite
often, roughly in 20\% of the cases.

{\it Two-level systems\/}.--- For a typical double-well potential
$E(\phi)$ with a given oscillator frequency $\omega$, the distance in
$\phi$ and the energy barrier between the two minima are of the order of
$\omega / \omega_p$ and $(c / R_c) (\omega / \omega_p)^4$, respectively.
Both decrease with $\omega$, and it can be verified that at $\omega \sim
\omega_q = K^{1/3} \omega_p$ where $K \equiv \hbar v / 2 \pi c \ll 1$,
the matrix element $I$ for the quantum tunneling between the two minima
becomes of the order of $\hbar \omega$. This implies that at frequencies
below $\omega_q$ the response of the chain is dominated by quantum
tunneling. It contributes to ${\Re\rm e}\,\sigma(\omega)$ whenever the
levels localized in the two minima are split in energy by exactly
$\hbar\omega$. A straightforward analysis~\cite{Il_in_87} of the model
defined by Eqs.~(\ref{L_eff}), (\ref{P_alpha_local_II}) and
(\ref{E_expansion}) leads to
\begin{equation}
{\Re\rm e}\,\sigma(\omega) \sim \frac{e^2}{h} R_c K^2
	\frac{\omega}{\omega_p},\quad
	\omega_{q1} \ll \omega \ll \omega_q.
\label{two_level_systems}
\end{equation}
The origin of the frequency scale $\omega_{q 1} \equiv \omega_p
\exp(-K^{-1})$ is as follows. The energy splitting of our two-level
systems (2LS) is bounded from below by $I$. Therefore, as $\omega$
decreases, the tunneling barrier and the tunneling distance have to
increase. At $\omega < \omega_{q 1}$ a typical 2LS cannot possess such a
small $I$ and Eq.~(\ref{two_level_systems}) ceases to be valid. As
$\omega$ continues to decrease, the dissipation is initially determined
by rare 2LS with unusually large barriers, until at $\omega_s = \omega_p
\exp(-K^{-3/2})$ the soliton mechanism becomes more prominent.

{\it Soliton tunneling\/}.--- In the soliton mechanism a large tunneling
action (small $I$) is due to a large tunneling mass. The object that
tunnels is a segment of the chain of length $l_{\rm tun}(\omega) \gg
R_c$. The optimal way to accomplish such a tunneling is to send a
virtual soliton (a $2 \pi$-kink in $\phi$) over the distance $l_{\rm
tun}$. Another route to the concept of solitons is appealing to
universality of the low-frequency physics of the glass phase (throughout
the range~\cite{Giamarchi_88} $K < 3/2$). If it holds, the conductivity
must be calculable by generalizing the Mott--Halperin
argument~\cite{Mott_68}. This argument devised for $K = 1$ focuses on
the tunneling of single electrons. But they are precisely the
$2\pi$-solitons in the bosonized formulation~(\ref{L}), so replacing the
word ``electron'' by ``soliton'' everywhere in the argument seems
entirely natural.

The 1D MHB law can be written in the form~\cite{Shklovskii_81}
\begin{equation}
{\Re\rm e}\,\sigma(\omega) \sim Q^2 l_{\rm tun}^2 l_s
                              \rho_s^2(\hbar \omega),
\label{MHB_3}
\end{equation}
where $\rho_s(E)$ is the density of states, $l_{\rm tun} \sim l_s \ln
(I_0 / \hbar \omega)$ is the typical tunneling distance, and $l_s$ is
the localization length of the tunneling charge-$Q$ objects ($Q = e$ for
$2\pi$-solitons). A typical soliton has the size $R_c$ and the
creation energy $E_s = c / R_c$, which implies $l_s \sim
\hbar v / E_s \sim K R_c$~\cite{Larkin_78}. Under the standard
assumption~\cite{Fisher_88} that $\rho_s(E) \to {\rm const} \sim (R_c
E_s)^{-1}$ for $E \to 0$, we get
\begin{equation}
{\Re\rm e}\,\sigma(\omega) \sim \frac{e^2}{h} R_c K^5
	\frac{\omega^2}{\omega_p^2}
	\ln^2\frac{\omega}{\omega_p},\quad \omega \ll \omega_s.
\label{MHB_2}
\end{equation}

\begin{figure}
\centerline{
\includegraphics[width=2.2in,bb=92 430 485 719]{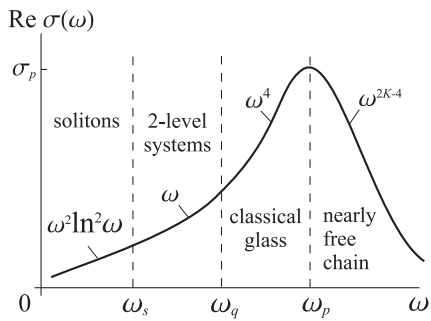}
}
\vspace{0.1in}
\setlength{\columnwidth}{3.2in}
\caption{
The sketch of ${\Re\rm e}\,\sigma(\omega)$. The nonperturbative
low-frequency regime $\omega < \omega_p$ is discussed in the text.
The formula for $\omega > \omega_p$ where the perturbation theory
applies is from \cite{Giamarchi_88}.
\label{Fig_sigma}
}
\end{figure}

The frequency dependence of the conductivity is summarized in
Fig.~\ref{Fig_sigma}. My results support the notion that the
infrared behavior of nondissipative classical~\cite{Il_in_87,Aleiner_94}
and quantum~\cite{Giamarchi_88} glasses are universally controlled by
the $\omega^4$-dependence~(\ref{classical_glass}) and the MHB
law~(\ref{MHB_2}), respectively.
In the model considered, as $K$ increases, the system becomes more
quantum. The quasiclassical regimes (2LS and $\omega^4$) shrink and
eventually get eliminated at $K \simeq 1$, where the soliton tunneling
crosses over directly to the Drude behavior, in agreement with
Ref.~\cite{Berezinskii_73}.

%
%

{\it Relation to experiments.\/}--- The obtained results have
implications for a wide variety of physical systems. Perhaps, the most
studied of them are the CDWs. Evidence for the 2LS in CDW compounds has
been seen in the specific-heat~\cite{Biljakovic_86}. Transport data
typically fit the $\sigma(\omega) \propto T \omega^b$ dependence with $b
\approx 1$~\cite{Wu_86}. It may or may not be due to the 2LS physics,
but in any case, it means the true zero-temperature limit I studied here
has not been realized in the experiment. To verify
Eqs.~(\ref{classical_glass}) and (\ref{two_level_systems}) new ac
transport experiments at considerably lower $T$ are needed. As for
Eq.~(\ref{MHB_2}), strong modifications due to Coulomb
effects~\cite{Shklovskii_81} are expected in real CDWs. Another class of
materials to test would be the high-spin spin-chain compounds whose
dynamics can be studied by electron spin resonance. Finally, some of the
ideas behind Eq.~(\ref{classical_glass}) may also be relevant for
understanding the $\omega^2$-broadening of phonons in glassy
liquids~\cite{Masciovecchio_96}.

{\it Acknowledgements.\/}--- I thank M.~Feigelman, T.~Giamarchi,
D.~Huse, O.~Motrunich, and V.~Vinokur for discussions and MIT
Pappalardo Program for support.


\begin{thebibliography}{99}
\bibitem{Comment_on_wavefunctions} Wavefunctions with spatial extent $l
\gg R_c$ may appear in special places where disorder is atypically
suppressed over the span of $\sim l / R_c$ elementary correlation
lengths $R_c$. Such more extended wavefunctions are exponentially rare
while their contribution to $\sigma(\omega)$ is enhanced merely by some
power of $l / R_c$. Therefore, they can be ignored.
\bibitem{Comment_on_s} Similar results were derived in
Refs.~\cite{Il_in_87} and \cite{Aleiner_94}.
\end{thebibliography}
\end{document}